\documentclass[doublecol,linenumbers]{epl2} 

\usepackage{dcolumn}    
\usepackage{bm} 
\usepackage{graphicx}
\usepackage{amsmath}    
\usepackage{latexsym}
\usepackage{amsfonts}   
\usepackage{amssymb}
\usepackage{array}      
\usepackage{epsfig}
\usepackage{dsfont}
\usepackage{color}
\usepackage[colorlinks=true,citecolor=blue,linkcolor=blue,urlcolor=blue]{hyperref}%

\newcommand{\bla}{bla\\bla\\bla\\bla\\bla}

\newcommand{\pket}[1]{\left\vert#1\right)}
\newcommand{\pbra}[1]{\left(#1\right\vert}
\newcommand{\pbraket}[1]{\left(#1\right)}
\newcommand{\st}[1]{\left\{#1\right\}}

\def\L{\mathcal{L}}

\begin{document}
\title{Operator growth and spread complexity in open quantum systems}

\author{Eoin Carolan$^{1,2}$, Anthony Kiely$^{1,2}$, Steve Campbell$^{1,2}$, and Sebastian Deffner$^{3,4}$}
\shortauthor{Eoin Carolan, Anthony Kiely, Steve Campbell, and Sebastian Deffner}

\institute{$^1$School of Physics, University College Dublin, Belfield, Dublin 4, Ireland \\
$^2$Centre for Quantum Engineering, Science, and Technology, University College Dublin, Belfield, Dublin 4, Ireland \\
$^3$Department of Physics, University of Maryland, Baltimore County, Baltimore, MD 21250, USA\\
$^4$National Quantum Laboratory, College Park, MD 20740, USA}

\abstract{ Commonly, the notion of ``quantum chaos'' refers to the fast scrambling of information throughout complex quantum systems undergoing unitary evolution. Motivated by the Krylov complexity and the operator growth hypothesis, we demonstrate that the entropy of the population distribution for an operator in time is a useful way to capture the complexity of the internal information dynamics of a system when subject to an environment and is, in principle, agnostic to the specific choice of operator basis. We demonstrate its effectiveness for the Sachdev–Ye–Kitaev (SYK) model, examining the dynamics of the system in both its Krylov basis and the basis of operator strings. We prove that the former basis minimises spread complexity while the latter is an eigenbasis for high dissipation. In both cases, we probe the long-time dynamics of the model and the phenomenological effects of decoherence on the complexity of the dynamics.}
\date{\today}
\maketitle

\section{Introduction} 

In classical mechanics, chaos is typically considered synonymous to an exponential sensitivity of the trajectories to initial conditions. However, rigorously defining quantum chaos is significantly more difficult \cite{Berry1989PS}. The first breakthrough for defining chaos in quantum systems originated in the development of random matrix theory, independently by Wigner \cite{wigner1993characteristic} and Dyson \cite{dyson1962brownian}. 

Subsequently, interest in dynamical signatures of chaos has grown, with much focus on the behaviour of autocorrelation functions \cite{kettemann1997characterization}, the Loschmidt echo \cite{chenu2018quantum}, and out-of-time-order correlation functions (OTOCs) \cite{
garcia2022out}. The latter, in particular, are notable as they place bounds on the rate of the spread of information in a system and can be used to calculate a quantum analogue of the Lyapunov exponent. However, the exponential decay of an OTOC is an indicator of \textit{scrambling}--the spread of initially local information throughout a many-body system--a necessary but not sufficient condition for a system to be chaotic \cite{xu2020does,dowling2023scrambling}. Much progress in recent years has come by shifting focus to the hydrodynamical behaviour of many-body quantum systems \cite{khemani2018operator,blake2018quantum,wienand2023emergence,rakovszky2022dissipation,rakovszky2018diffusive,von2018operator}, and the spread of support of both states and operators \cite{balasubramanian2022quantum,bhattacharya2023spread,caputa2024krylov,arpan2023}. 
While an intuitive picture for operator growth is viewing it as the spread of support of an operator in physics space, we stress that the \emph{growth of an operator} refers to the increasing support in Hilbert space, which is a more applicable definition to models that have non-local interactions.

Recent work has explored the competition and similarities of scrambling and decoherence \cite{arpan2023, touil2021information,zanardi2021information,larzul2022fast,mohan2023krylov,xu2019extreme,weinstein2023scrambling,tripathy2024chaos,garcia2024lyapunov}. Decoherence is a channel for information from the system to leak into the environment, as opposed to being spread into entanglement structures in a many-body system. The OTOC has been shown to not distinguish between these two effects \cite{touil2021information}, meaning that other measures must be used in the open system setting.

Complex spacing ratios \cite{Prosen2020} and dissipative form factors \cite{li2021spectral,arpan2023} have been developed to characterise the level repulsion for chaotic systems in non-Hermitian settings, where the spectrum is no longer purely real. The operator growth hypothesis (OGH) \cite{Parker_2019} has gained significance as a tractable method of calculating the complexity of a system and has placed upper bounds on the Lyapunov exponent extracted from the OTOC at infinite temperature. While successful for closed systems and Markovian master equations, it is not clear if the methods prescribed by the OGH can be applied to more general maps such as those that give rise to non-Markovian evolution. 

This provides the starting point of our work, in which we propose a tractable method for characterising chaotic dynamics and operator complexity in open quantum systems. In particular, we introduce the \emph{operator spread entropy} as a general notion for examining operator growth that provides a measure of complexity as well as allowing for insights into the operator population dynamics. Crucially, the spread entropy does not prescribe a particular basis, and therefore in principle any suitable basis can capture the same qualitative behaviour for the spread entropy as the Krylov basis. We apply our framework to the Sachdev-Ye-Kitaev (SYK) model, probing the late time dynamics under decoherence, complementing recent work on operator complexity and decoherence~ \cite{schuster2022operator,liu2024operator}.

\section{Operator Growth and the Bi-Lanczos Algorithm}

The OGH \cite{Parker_2019} defines an analogue of the classical Lyapunov exponent even for quantum systems which may not have a well-defined semi-classical limit. To determine the growth rate, consider a system described by a Hamiltonian $H$, and an initially local Hermitian operator, $X_0$, which one may view as a ``vector" in the operator Hilbert space, denoted by $\pket{X_0}$. The operator evolves under the action of the superoperator $\mathcal{L}$. 

The Maclaurin series expansion of the operator follows as
$  \pket{X_t} = e^{i \mathcal{L} t} \pket{X_0} =\sum_n \frac{(it)^n}{n!}\mathcal{L}^n \pket{X_0}$.
In the closed case $\mathcal{L} \bullet:= [H,\bullet]$, and operators that do not correspond to a conserved quantity with respect to the Hamiltonian will spread in support with repeated applications of this commutator. All of the information about the evolution of the operator is therefore contained in the set $\{\mathcal{L}^n \pket{X_0}\}$, which is intuitively the minimal basis needed to encode the dynamics. 

We will focus on the case when the time evolution of an operator $X_0$ is governed by a Markovian Lindblad master equation,
\begin{equation}\label{eq:gksl}
    \begin{split}
    \frac{dX_0}{dt} &= i\mathcal{L} X_0, \\
    &=i [H,X_0]+\sum_n \mu_n[\pm L_n^{\dagger}X_0 L_n-\frac{1}{2}\{L_n^{\dagger}L_n,X_0\}],
    \end{split}
\end{equation}
where $``-"$  is taken when both the operator $X_0$ and the jump operators $L_n$ are fermionic~\cite{liu2023krylov} and $``+"$ otherwise. To create a basis for the dynamics from $\{\mathcal{L}^n \pket{X_0}\}$ we first need to define an inner product. Due to an inherent ambiguity over what inner product one should take at finite temperatures~\cite{caputa2022geometry,Parker_2019}, we choose to take the infinite-temperature inner product $\left(A\vert B\right) := \text{Tr}\left[A^\dagger B\right] / \text{Tr}\left[\mathds{1}\right]$

A variety of methods to create an orthonormal basis in this setting have been explored~\cite{bhattacharya2023krylov,srivatsa2023operator,bhattacharjee2023operator,bhattacharya2022operator,liu2023krylov}. We shall focus on the bi-Lanczos algorithm \cite{gaaf2017infinite,bhattacharya2023krylov}, which satisfies the conditions for being a ``\textit{K}-complexity" as defined in~\cite{Parker_2019}, and recovers the Lanczos algorithm for zero decoherence. 

The bi-Lanczos algorithm evolves the left and right vectors of $X_0$ separately, enforcing orthonormality between elements of each set. We first set $b_0=c_0=0$, and then proceed with the bi-Lanczos algorithm,
\begin{equation}
	\begin{aligned}
	\pket{A_n} &:=(\mathcal{L}-a_{n-1} ) \pket{ O_{n-1}} - c_{n-1} \pket{ O_{n-2}} \,, \\
    \pket{B_n} &:=(\mathcal{L}^{\dagger}-a^*_{n-1} ) \pket{ O_{n-1}} - b_{n-1} \pket{ O_{n-2}}, \\
	\pket{O_{n}} &:= \ b_{n}^{-1} \pket{A_n} \,,\;\;  \pbra{\tilde{O}_{n}} =  c_{n}^{-1} \pbra{B_n},~~~~~~~~\text{with} \\
    a_n &:= \pbraket{\Tilde{O}_n \vert \mathcal{L} \vert O_n} ,\;   b_n = \sqrt{\pbraket{A_n \vert A_n}} ,\; c_n = \frac{\sqrt{\pbraket{B_n \vert A_n}}}{b_n} .
	\end{aligned}
	\label{eq:biLanczos_algorithm}
	\end{equation}
The algorithm terminates when $b_n\!=\!0$ for finite systems or when successive Krylov basis elements align. This termination condition involves a numerical tolerance. We therefore remark that the dimension of the Krylov space calculated reflects the number of numerically relevant elements, which may not be the exact dimension of Krylov space. We output two sets of vectors for which we have the orthogonality relation $\pbraket{\Tilde{O}_n|O_m}=\delta_{nm}$, where we remark that each set by itself is not necessarily orthogonal and in the bi-Lanczos basis the superoperator takes the tri-diagonal form
 \begin{equation}
	\mathcal{L} = \sum_{n,m} \pbraket{\tilde{O}_n | \mathcal{L} | O_m} \pket{O_n} (\tilde{O}_m |  = \begin{pmatrix}
	a_0 & b_1 & 0 & \cdots\\
	c_1 & a_1 & b_2  & \cdots\\
	0 & c_2 & a_2  & \cdots\\[-0.3em]
	\vdots & \vdots & \vdots & \ddots
	\end{pmatrix}.
	\label{eq:L_tridiagonal}
	\end{equation}
 
The superoperator is analogous to the tight-binding chain, which we will refer to as the ``Krylov chain". In this picture, the operator ``hops" to higher basis elements in the Krylov chain over time, with higher-$n$ elements generically having larger support. We can write the time-evolved operator in both spaces as
\begin{equation}\label{eq:opk}
    \pket{X_t}=\sum_{n=0}i^n\phi_n(t)\pket{O_n},~
    \pbra{X_t}=\sum_{n=0}(-i)^n\varphi_n^*(t)(\tilde{O}_n\vert.
\end{equation}
The Krylov basis allows us to define a dynamical indicator of scrambling, the Krylov complexity \cite{Parker_2019,bhattacharya2023krylov}
\begin{equation}
\label{eq:kComplexity}
K(t)=\frac{\sum_{n=0}^{M_{\mathcal{K}}-1}n\varphi_n^*(t)\phi_n(t)}{\sum_{n=0}^{M_{\mathcal{K}}-1}\varphi_n^*(t)\phi_n(t)},
\end{equation}
where $M_{\mathcal{K}}$ is the dimension of Krylov space. Due to the fact that the norm of the operator is not preserved in time for open dynamics, we have to renormalise the populations.
We can interpret eq.~\eqref{eq:kComplexity} as the expected position on the Krylov chain that the operator lies on i.e. how deeply it has saturated into the Krylov basis. Closer study of the Krylov complexity has given additional insight into operator growth \cite{hornedal2022ultimate,hornedal2023geometric}, deriving the conditions needed for a model to saturate an upper bound on its rate of change. The SYK model is such a system. The methods of Krylov complexity have also been extended to the Trotter decomposition of unitary dynamics \cite{suchsland2023krylov}, allowing for new tools to study phenomena in quantum circuits.

We note that the $a_n$ coefficients are purely imaginary, while the $b_n$ and $c_n$ are purely real. In the limit of closed system dynamics $\mu\to0$, we have  $a_n\to0$, $c_n \rightarrow b_n$ and $\varphi_n \rightarrow \phi_n$, recovering both the Krylov basis and the Krylov complexity of the closed dynamics which forms the foundation of the OGH \cite{Parker_2019}.

In the closed case, the OGH states that the asymptotic growth of the Lanczos coefficients is maximal for chaotic systems. Specifically, this is characterised by a linear rate, $\alpha>0$, such that  $b_n\!=\!\alpha n+\gamma$ where $\gamma$ is a constant. The OGH has been successful in demonstrating the linear growth of Lanczos coefficients for chaotic systems, both analytically and numerically, for a number of models~\cite{doublescaled,bhattacharjee2022probing,rabinovici2022krylov,noh2021operator}. However, while chaotic systems exhibit a linear growth in the Lanczos coefficients, unstable yet integrable systems may do the same~\cite{bhattacharjee2022krylov,huh2023spread}. Thus, chaos implies a linear growth in the Lanczos coefficients, but the converse does not necessarily follow.

The Krylov basis output by the bi-Lanczos algorithm is the minimal basis for describing the open dynamics of a particular operator, making it the natural choice of basis from which to extract universal behaviour. We use it to probe the long time dynamics of the SYK model under decoherence, and demonstrate its usefulness as a basis for the operator spread complexity. While the bi-Lanczos algorithm lets us probe operator complexity for dynamics generated by a Markovian Lindbladian, it remains to be seen how one can generate the Krylov basis for general open system dynamics, or even those for which the superoperator is not accessible such as a collision model~\cite{csenyacsa2022entropy, li2022dissipation}.

\section{Operator Spread Complexity}
We now define an operator complexity measure for general dynamics. Consider a general orthonormal Hermitian operator basis $\mathcal{G}=\st{\pket{G_n}}$. The normalised overlap of an operator, $\pket{X_t}$, at a time $t$ with the $n^{\rm th}$ element of this basis is
\begin{eqnarray}\label{eq:p_b}
P_{\mathcal{G}}(n,t) &=& \frac{ \left|\pbraket{G_n\vert X_t}\right|^2}{\sum_m \left|\pbraket{G_m\vert X_t}\right|^2}.
\end{eqnarray}

The population distribution of an operator can be used to study the onset of quantum chaos and has been shown to be intrinsically related to the OTOC \cite{omanakuttan2023scrambling,blocher2023probing}. It is explicitly applied to the Krylov space for closed dynamics in \cite{barbon2019evolution}. To turn it into a measure of complexity we first demand that $\pket{G_0}=\pket{X_0}$.
The extent of the operator in a given basis \cite{campbell1966exponential} is given by the complexity (which can be recognised as both the diversity \cite{jost2006entropy} and perplexity \cite{jelinek1977perplexity} of a distribution)
\begin{equation}
\label{eq:ComplexityEntropy}
C_{\mathcal{G}}(t)=e^{F_{\mathcal{G}}(t)},
\end{equation}
where $F_{\mathcal{G}}(t)=-\sum_n P_{\mathcal{G}}(n,t) \ln P_{\mathcal{G}}(n,t)$ is the Shannon entropy for the operator distribution. For $t=0$ we have $C_{\mathcal{G}}(0)=1$, which increases with time due to scrambling of the operator, ultimately saturating at long times if the operator is maximally spread over all available basis elements. 

Two things must be noted about this measure: Firstly, we must choose {\it a-priori} a basis to measure the spread of the system over. Secondly, this measure distinguishes average operator size and complexity. For instance, consider a spin chain system described using a basis constructed from the strings of Pauli matrices. One could imagine a scenario where the time evolved operator has full {\it spatial} support over the chain, but is nevertheless ``simply" a linear combination of a few strings of this maximal length. The spread complexity will be low in this case, thus reflecting its low complexity in the bulk of the spin chain. This may be the case for certain Clifford circuits \cite{fisher2023random}. It is therefore relevant to consider whether other bases, aside from the minimal one, capture operator dynamics accurately.

\section{Minimisation of the spread complexity}
To show that the Krylov basis minimises the operator spread complexity we will utilise a similar approach as that given for the spread complexity of a state governed by the Schr{\"o}dinger equation, derived in ref. \cite{balasubramanian2022quantum}. We modify their starting point to that of the evolution of an operator governed by a superoperator, with the only caveat being that we can obtain the Krylov space for the dynamics. Taking $k$ derivatives of eq.~\eqref{eq:p_b} gives
\begin{eqnarray}\label{eq:pder}
    &&P_{\mathcal{G}}^{(k)}(n,t)=\frac{\partial^kP_{\mathcal{G}}(n,t)}{\partial^kt}\nonumber\\
    &=&\frac{\sum_{j=0}^ki^k(-1)^{j}\binom{k}{j}\pbra{X_t}\mathcal{L}^{\dagger j}\pket{G_n}\pbra{G_n}\mathcal{L}^{k-j}\pket{X_t}}{\|X_t\|^2}\\
    &+&\frac{\pbraket{X_t|G_n}\pbraket{G_n|X_t}\partial_t\|X_t\|^2}{\|X_t\|^4},
\end{eqnarray}
where we recognise that $\sum_n \left|\pbraket{G_n\vert X_t}\right|^2=\|X_t\|^2$ for a complete basis.
Let us assume for both a general basis, $\mathcal{G}$, and the Krylov basis $\{\pket{O_n}\}\in\mathcal{K}$ that the first element, i.e, $n=0$, is $X_0$, and that the following $m-1$ elements are common to both. Therefore for $n<m$, we have that $P_{\mathcal{K}}^{(k)}(n,t)=P_{\mathcal{G}}^{(k)}(n,t)$. 

\noindent
{\bf Lemma} {\it If the first $m$ elements of $\mathcal{G}$ are those of $\mathcal{K}$ then $P_{\mathcal{G}}^{(k)}(n,0)=0$ for $n\geq m$ and $k<2m$.}

\noindent{\it Proof.}  From eq.~\eqref{eq:pder} we see that $P_{\mathcal{G}}^{(k)}(n,0)$ has at most $k$ applications of the superoperator to $\pket{X_0}$. 
Taking $k<m$, it is clear that $\pbra{G_n}\mathcal{L}^{k}\pket{X_0}=\pbra{X_0}\mathcal{L}^{\dagger k}\pket{G_n}=0$ for $n\geq m$ and $k<m$ as $\mathcal{L}^{k}\pket{X_0}$ requires at least $k=m$ applications of $\mathcal{L}$ to generate overlap with the first element of $\mathcal{G}$ that is not also a Krylov element, $\pket{G_m}$. We note that for all $n>0$, the final term in eq.~\eqref{eq:pder} is zero at $t=0$ as $\pbraket{G_n|X_0}=\delta_{n,0}$. For $k<2m$, all of the terms in the sum for $P_{\mathcal{G}}^{(k)}(n,0)$ will be zero as either $\pbra{X_0}\mathcal{L}^{\dagger j}\pket{G_n}$ or $\pbra{G_n}\mathcal{L}^{k-j}\pket{X_0}$ will involve less than $m$ applications of the superoperator to $\pket{X_0}$, making it zero by the same argument, proving the lemma. $\square$

The spread complexity differs between the two cases when $n\geq m$. We write the Shannon entropy of the terms that differ from the Krylov basis as
\begin{equation}
    F_{n\geq m}(t)=-\sum_{n\geq m} P_{\mathcal{G}}(n,t)\ln  P_{\mathcal{G}}(n,t).
\end{equation}
We are interested in the behaviour of  $P_{\mathcal{G}}(n,t)$ when $n \geq m$. We invoke the lemma to identify that the first non-zero term in the Taylor series expansion around $t=0$ occurs when $k=2m$,
\begin{eqnarray}
     P_{\mathcal{G}}(n,t) &=& \sum_k\frac{P_{\mathcal{G}}^{(k)}(n,0)t^k}{k!} \nonumber \\
     &=& \frac{P_{\mathcal{G}}^{(2m)}(n,0)t^{2m}}{(2m)!}+\mathcal{O}(t^{2m+1}).     \label{eq:expansion}
\end{eqnarray}
We substitute in eq.~\eqref{eq:expansion}, and split the logarithm term into two separate parts,
\begin{eqnarray}\label{app:eq_split}
    F_{n\geq m}(t)&=&-\frac{\ln(t)t^{2m}}{(2m-1)!}\sum_{n\geq m}P^{(2m)}_{\mathcal{G}}(n,0)\\
     &-&\sum_{n\geq m}\frac{P^{(2m)}_{\mathcal{G}}(n,0)t^{2m}}{(2m)!}\ln\big[P^{(2m)}_{\mathcal{G}}(n,0)/(2m)!\big]. \nonumber
\end{eqnarray}
The non-zero part of $P^{(2m)}_{\mathcal{G}}(n,0)$ can be written as
\begin{equation}
    P^{(2m)}_{\mathcal{G}}(n,0)=\binom{2m}{m}\pbra{X_0}\mathcal{L}^{\dagger m}\pket{G_n}\pbra{G_n}\mathcal{L}^{m}\pket{X_0},
\end{equation}
noting that $\|X_0\|=1$ and that the final term in eq.~\eqref{eq:pder} is zero for all $n>0$. The non-zero contribution here comes from $\pket{Y}$, which is the part of $\mathcal{L}^m\pket{X_0}$ orthogonal to the first $m$ basis elements. 
We then write
\begin{equation}
    \sum_{n\geq m}P^{(2m)}_{\mathcal{G}}(n,0)= \sum_{n\geq m}\binom{2m}{m}\pbraket{Y|G_n}\pbraket{G_n|Y}.
\end{equation}
As $\pket{Y}$ is orthogonal to the first $m$ elements of the basis, we can extend this sum to start at zero, and invoke the completeness of $\mathcal{G}$ to write
\begin{equation}
     \sum_{n\geq m}P^{(2m)}_{\mathcal{G}}(n,0)=\binom{2m}{m}\pbraket{Y|Y},
\end{equation}
which greatly simplifies the first part in eq.~\eqref{app:eq_split} into something that is basis independent. The second term has the form $f(x)=\frac{x}{(2m)!}\ln\frac{x}{(2m)!}$ with $x=P^{(2m)}_{\mathcal{G}}(n,0)$, which is a convex function that is negative for the domain considered. $P^{(2m)}_{\mathcal{G}}(n,0)$ is a positive number for which the sequence $(\alpha_i)=\left(\binom{2m}{m}\pbraket{Y|Y},0,0,...0\right)$ trivially majorises any other sequence $(\beta_i)$ of positive numbers that add to $\binom{2m}{m}\pbraket{Y|Y}$. This implies, by Karamata's inequality \cite{kadelburg2005inequalities}, that $\sum f(\alpha_i)\geq \sum f(\beta_i)$, i.e., that the entropy (once we take the overall minus sign) is always greater than or equal to the case where $\sum_{n\geq m}P^{(2m)}_{\mathcal{G}}(n,0)$ has only contribution from a single basis element, meaning that the Krylov basis element $\pket{O_m}$ must be part of the basis to minimise this term, allowing us (by induction) to conclude that the Krylov basis minimises the entropy for the population distribution for both closed evolution and under dynamics generated by a Markovian Lindbladian. 

\section{Sachdev-Ye-Kitaev Model}
To demonstrate our framework, we analyze the SYK model, which consists of $N$ interacting Majorana fermions. This system is a paradigmatic model of quantum chaos~\cite{Sachdev_1993}. Majorana fermions, $\psi_i$, are defined through their anticommutation relation
    $\{\psi_i,\psi_j\}=\delta_{ij}$
and the dimension of the Hilbert space of $N$ Majorana fermions is $2^{N/2}$. The SYK model is an all-to-all coupled model with the Hamiltonian
\begin{equation}
    H_{SYK}=(i)^{q/2}\sum_{1\leq i_1< i_2<...<i_q\leq N}J_{i_1i_2...i_q}\psi_{i_1}\psi_{i_2}...\psi_{i_q},
\end{equation}
where $q$ denotes the number of fermions that interact in a vertex, $q=2$ being an integrable free fermion model, and $q>2$ giving rise to chaotic behaviour. The sum is ordered in such a way as to include interactions between any $q$ fermions once, and the interaction strength is a real number $J_{i_1i_2...i_q}$ drawn from a random Gaussian distribution with a zero mean and a variance
\begin{equation}
    \overline{J^2_{i_1i_2...i_q}}=\frac{J^2(q-1)!}{N^{q-1}},
\end{equation}
where the overline denotes the disorder average.

The SYK model is both a maximally chaotic model (viewed through the framework of the operator growth hypothesis \cite{Parker_2019}) and a fast scrambler \cite{sekino2008fast,belyansky2020minimal}). Other models exhibit this behaviour, such as random unitary circuits~\cite{belyansky2020minimal,sekino2008fast}. Importantly, it saturates the bound on the rate of change of Krylov complexity \cite{hornedal2022ultimate,hornedal2023geometric}.

While the SYK and its symmetries are well understood as a closed many-body system, only recently has its open dynamical behaviour been examined~\cite{Xu_2021,Kulkarni_2022,Ryu2022a,ryu2022b,sa2022lindbladian,syk2024}.
\begin{figure*}[t]
\begin{center}
{\bf (a)} \hskip0.5\columnwidth {\bf (b)} \hskip0.5\columnwidth {\bf (c)}\\
\includegraphics[width=0.55\columnwidth]{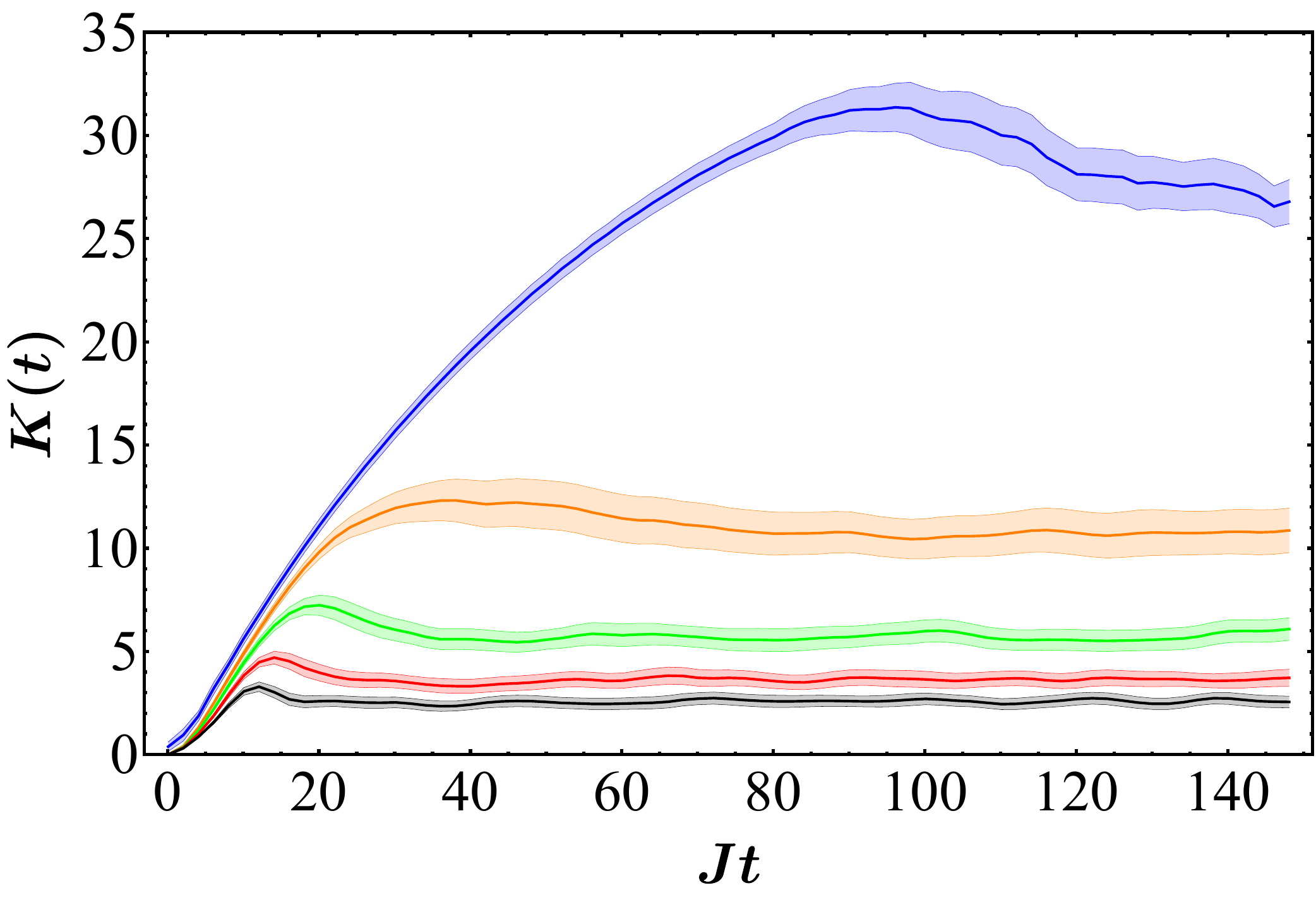}
~\includegraphics[width=0.55\columnwidth]{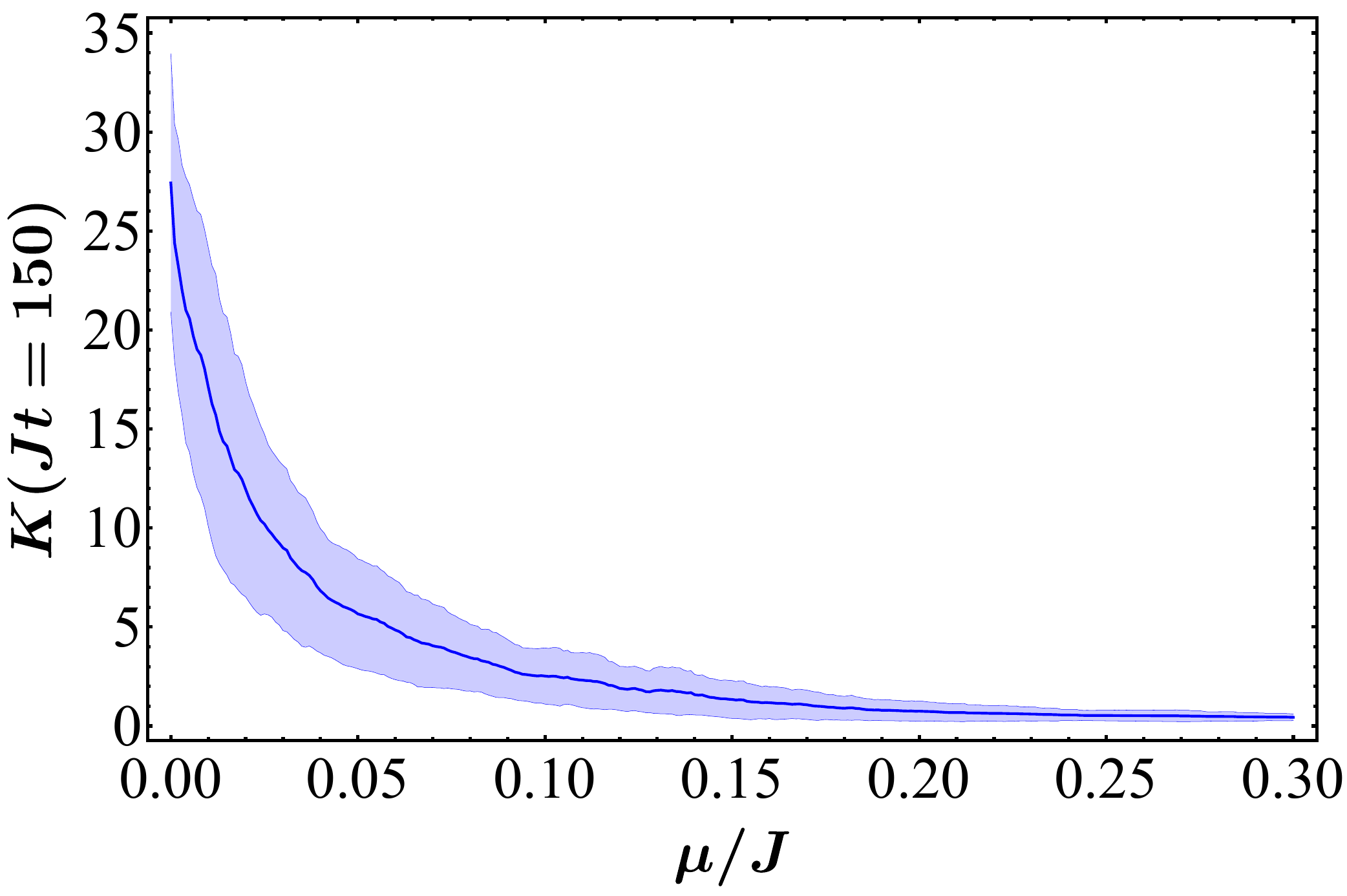}~\includegraphics[width=0.55\columnwidth]{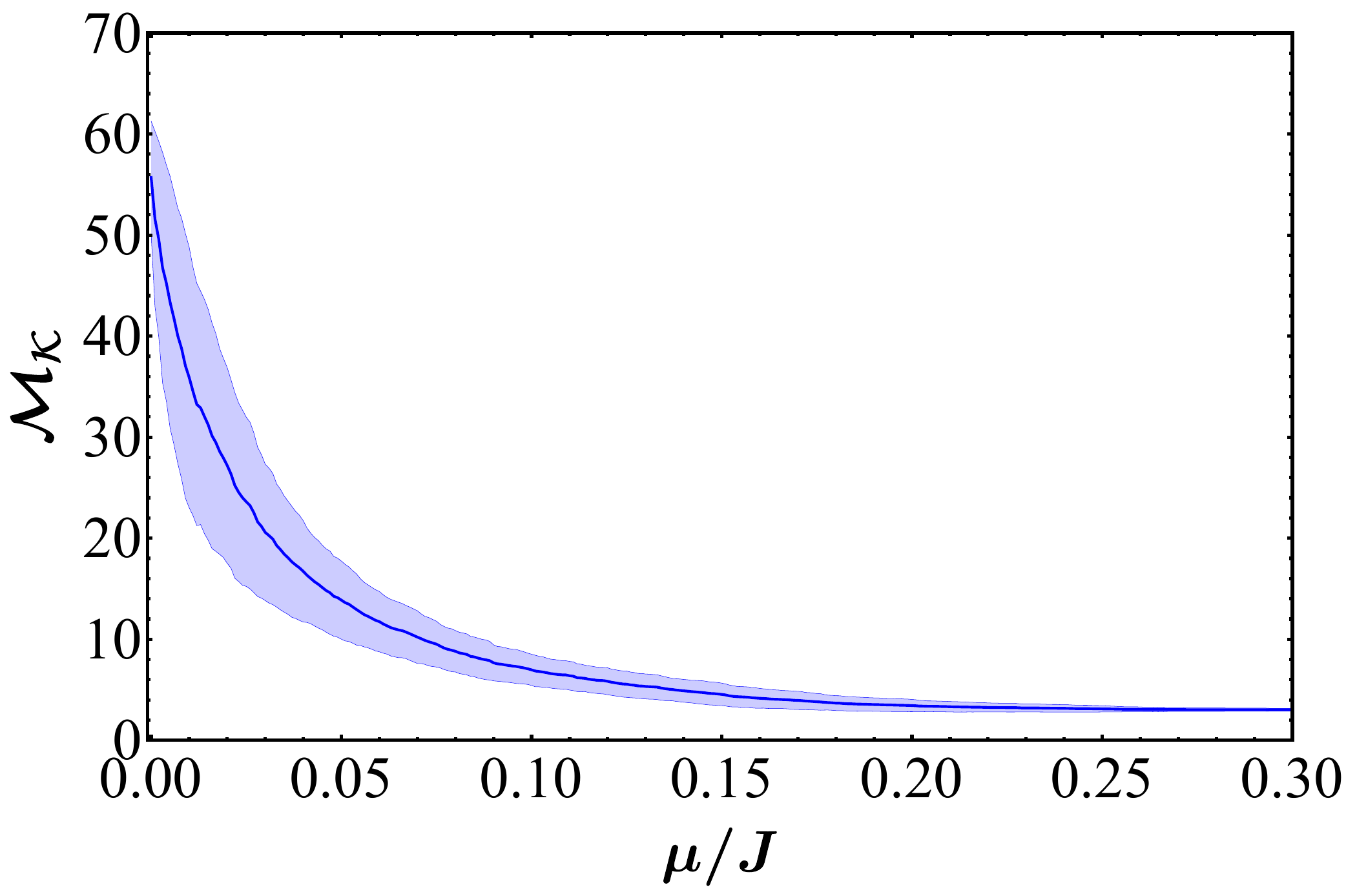}
\end{center}
\caption{ Dynamics for the open SYK model with $N=8$ and 200 disorder realisations. Variance is shown as a shaded region (scaled to 20\% of its value in panel (a) for clarity). (a) The average Krylov complexity (eq.~\eqref{eq:kComplexity}) with $\mu/J=0.0, 0.025, 0.05, 0.075, 0.1$ corresponding to blue, orange, green, red, and black, respectively. {\bf (b)} The average Krylov complexity at $Jt=120$ vs decoherence strength $\mu/J$.   {\bf (c)} The average dimension of the Krylov space for the dynamics vs $\mu/J$.}
\label{figcomp}
\end{figure*}

\section{Operator growth and spread complexity for the open SYK model}
We consider a Markovian Lindblad master equation for the SYK model as given by eq.~\eqref{eq:gksl} with the minus sign taken. The fermionic jump operators are $L_n=\sqrt{\mu}\psi_n$ where $\mu$ governs the strength of the dissipation. We write it as $\mathcal{L} = \mathcal{L}_U+\mathcal{L}_D$ where
\begin{eqnarray}
    \mathcal{L}_U\bullet &=& [H_{SYK},\bullet] , \\
    \mathcal{L}_D\bullet &=& i\mu\sum_{n=1}^{N}\Big(\psi_n\bullet\psi_n+\frac{1}{4}\big\{\mathds{1},\bullet\big\}\Big),
\end{eqnarray}
and we have used the anticommutation relation of the Majorana operators and that they are Hermitian.

As we shall discuss, the action of the dissipative part of the master equation leads to a dampening of the Majorana string terms contributing to the time evolved operator at a rate proportional to their size. We define a Majorana string $S_i$ of length $s_i$ to be an operator formed as a product of $s_i$ Majorana fermions ordered such that the indices are in ascending order from left to right, e.g., $\psi_1\psi_3\psi_7$ is a string of length three. We will use the set of Majorana strings as an orthonormal basis for the spread complexity of the SYK model. For the $q\!=\!4$ SYK model under dissipation, we only need half of the complete basis since only strings of Majorana fermions of odd length can be generated by the interaction vertices provided the initial operator has odd length. Unlike the Krylov basis for the SYK model, this basis is fixed and identical for each iteration for the SYK model. As a basis it is physically well-motivated as it can be directly used to track the size of operators \cite{roberts2018operator}. 

The precise behaviour of the decoherence channel we apply to the SYK model has been established in \cite{bhattacharjee2023operator}. We will interpret these results to show that decoherence decreases complexity and Krylov space dimension. We assume an initial operator $X_0=\sum_i p_i S_i$. 

Considering the action of the non-unitary, decohering term on one of the strings for now, we find
\begin{equation}
    \mathcal{L}_D S_i=i\mu \sum_{n=1}^N\psi_n S_i\psi_n+i\mu \frac{N}{2} S_i.
\end{equation}
Now, we anticommute the first $\psi_n$ through $S_i$ which will allow us to then square it to $\mathds{1}/2$. This typically takes $s_i$ anticommutations to move it through, unless $\psi_n$ appears in the string $S_i$, in which case it takes $s_i-1$,
leaving
\begin{eqnarray}
       \mathcal{L}_D S_i= \frac{i \mu}{2} \left[ (-1)^{s_i}(N-s_i)+(-1)^{s_i-1} s_i+N\right] S_i. 
\end{eqnarray}

Depending on whether $s_i$ is odd or even, the right-hand side reads $i\mu s_i S_i$ or $i\mu (N-s_i) S_i$ respectively. Once we apply the same process to each of the strings that appear in a linear combination to make $X_0$, we obtain an operator that is co-linear with $X_0$
\begin{equation}
\mathcal{L}_D X_0=i\mu\sum_i\alpha_i p_i S_i,
\end{equation}
where $\alpha_i=s_i$ or $N-s_i$ depending on whether $s_i$ is odd or even. Note that for our purposes, only odd strings are relevant for the dynamics, meaning that the decoherence term dampens strings at a rate proportional to their length. It is clear that the unitary part is the source of the new operators appearing in the support of the time evolved operator to Majorana strings not originally present in $X_0$, as described in \cite{roberts2018operator}. In the limit of strong decoherence, the Lindbladian does not generate any new support from its action on the initial operator, so viewed through the lens of the Lanczos algorithm, it terminates immediately, giving $M_{\mathcal{K}}=1$. 

This has similarities to ref.~\cite{schuster2022operator}, where the model of decoherence acts like a measurement operator that is sensitive to the string length. The interplay between information scrambling and decoherence interpolates between the closed case, where the system generates as much support as is available to it, and the ``Zeno-blocked" case where decoherence term is measuring the initial string sufficiently strongly such that it does not grow in support. In the limit of strong decoherence, the operator evolves as
\begin{equation}\label{eq:eigenoperators}
    X_t\approx \sum_i p_ie^{-\mu s_i t}S_i.
\end{equation}
Thus, in the limit of strong decoherence, we see that any operator strings are eigenoperators for the Lindbladian.

We next compare how the two bases--the Krylov basis $\mathcal{K}$ generated from applying the bi-Lanczos algorithm and the Majorana string basis $\mathcal{S}$ ~\cite{roberts2018operator,schuster2022operator}--capture the spread complexity. Clearly the Krylov basis is the natural choice to examine universal behaviour and growth rates for systems. However, the latter is arguably a more natural basis for understanding the dynamics explicitly in terms of length of operator size. We fix $\sqrt{2}\psi_1$ as the initial operator.

\begin{figure}
\begin{center}
{\bf (a)}\\
\includegraphics[width=0.6\linewidth]{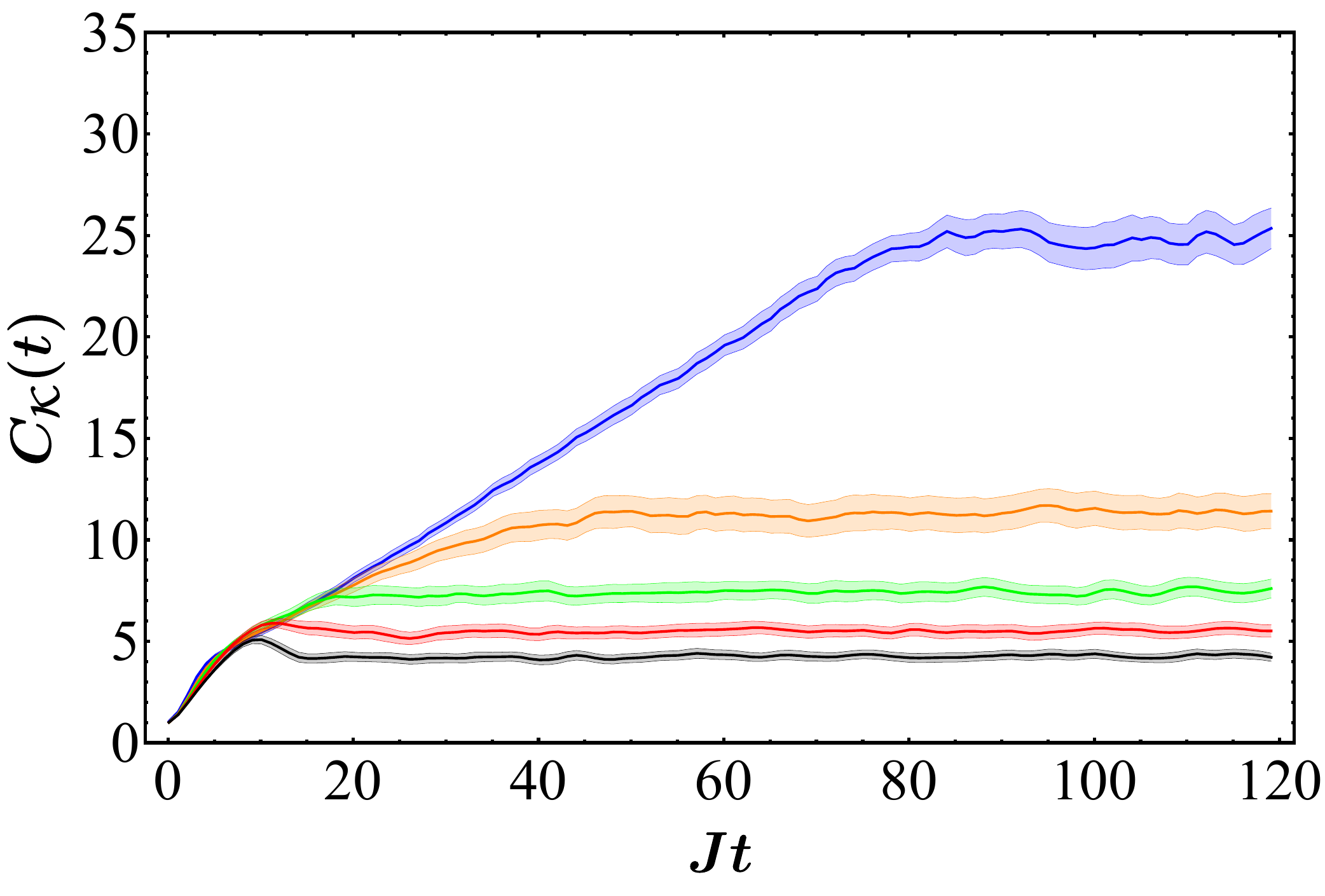}\\
{\bf (b)}\\
\includegraphics[width=0.6\linewidth]{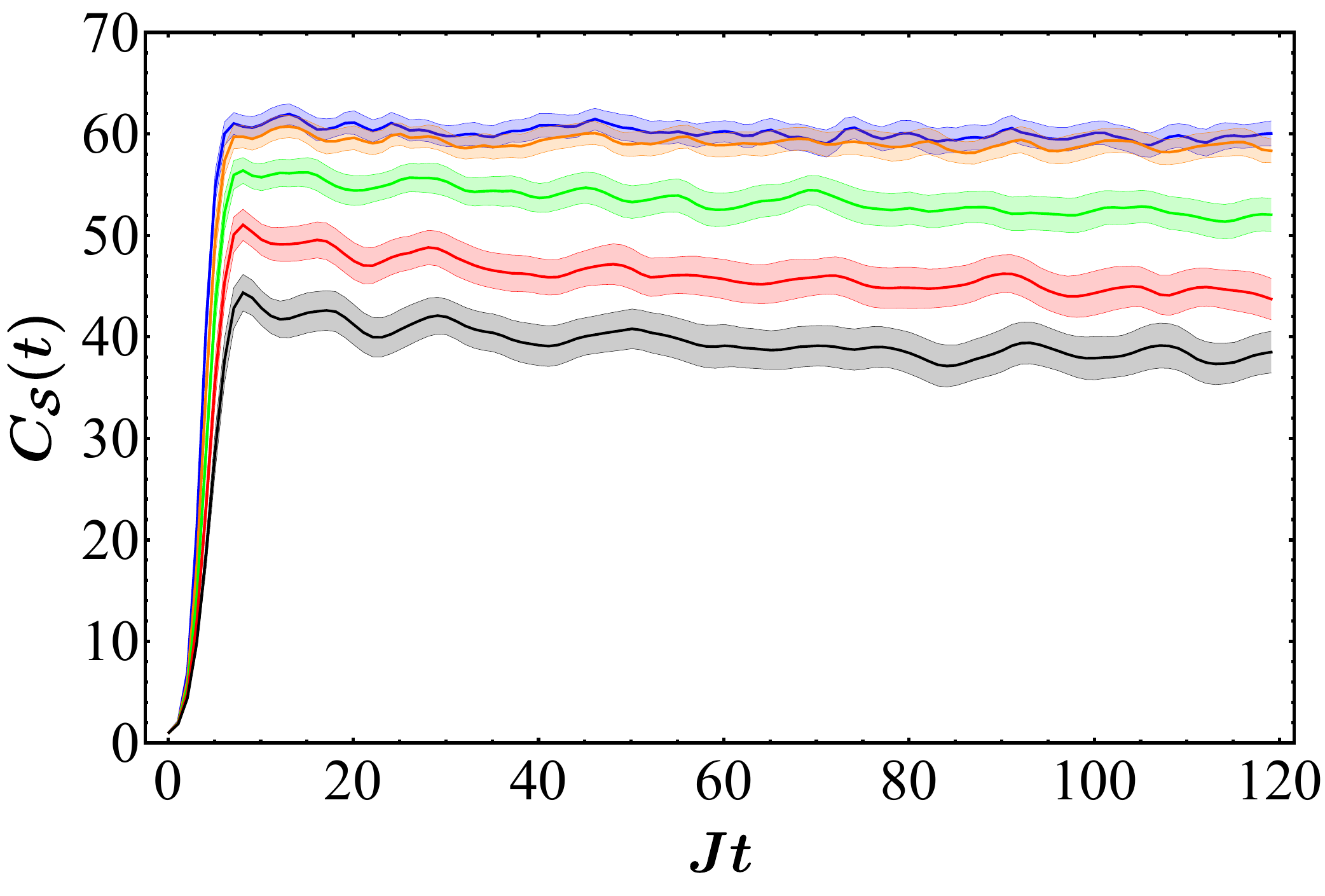}
\end{center}
\caption{ Dynamics for the open SYK model with $N=8$ and 200 disorder realisations with $\mu/J=0.0, 0.025, 0.05, 0.075, 0.1$ corresponding to blue, orange, green, red, and black, respectively. Variance is shown as a shaded region scaled to 20 \% of its value for clarity. {\bf (a)} Average spread complexity (eq.~\eqref{eq:ComplexityEntropy}) in the Krylov basis. {\bf (b)} Average spread complexity in the string basis.}
\label{figent}
\end{figure}

Figure \ref{figcomp}(a) depicts the Krylov complexity for the open SYK model over $200$ disorder realisations for a range of dissipation strengths. We also show the closed case (topmost blue line), i.e., $\mu=0$, where the bi-Lanczos algorithm reduces to the regular Lanczos algorithm. We see that initial growth in the closed case and under weak decoherence is similar, however they saturate at different levels. This behaviour is consistent with the large-N behaviour of the SYK Lindbladian model as established in \cite{syk2024}, where the Krylov complexity is shown to plateau at smaller levels for increasing dissipation strength. The saturation level of the Krylov complexity for a range of decoherence strengths is shown in fig. \ref{figcomp}(b). Its decreasing value as the open system effects become stronger indicates that information scrambles less throughout the system when subject to decoherence and the dynamics become less ``complex". Why the complexity of the dynamics is reduced under decoherence becomes clear when we plot the dimension of the Krylov basis for the SYK Lindbladian vs decoherence strength in fig. \ref{figcomp}(c). The dimension of the Krylov space corresponds to the number of elements needed to encode $\{\L^n\pket{X_0}\}$. Even in the closed case we see that the Lanczos algorithm compresses this information down into fewer basis states than needed for the entire Hilbert space. As it becomes less likely for our operator to inhabit regions of Krylov space with increasing decoherence strength, this set can be compressed down further. The scaling of the complexity naturally corresponds to the scaling of the Krylov space. This suggests a competition between information loss to the environment and the ability for a system to scramble its information internally. 

Somewhat naturally, the cardinality of the Krylov basis appears as \emph{the} quantity to infer the scrambling nature of a system. However, only the Krylov complexity, which weighs the contribution of the basis elements, is a genuninely dynamical quantity from which scrambling times and growth rates can be derived. Hence, we plot the operator spread complexity (eq.~\eqref{eq:ComplexityEntropy}) vs time in fig. \ref{figent}. Both the Krylov and string bases show the same hierarchy in spread complexity for different decoherence strengths. The rapid early growth of spread complexity in the string basis case comes from the inherent non-local nature of the SYK model. A few applications of the superoperator is all that is needed to have contributions from all strings in the basis. We postulate that for a local model, the qualitative growth of the spread complexity in both the Krylov and (Majorana or Pauli) string bases should be closer. This opens the door to moving past Markovian dynamics, allowing to assess whether information back flow into the system has a potential competing effect alongside decoherence and internal scrambling. Maps that generate dynamics with information back flow, even if they can be written in a master equation form, are not amenable to the bi-Lanczos approach. A pre-chosen basis, such as the string basis, removes this roadblock. Since this basis still allows to accurately capture the correct qualitative behavior as evidenced from fig.~\ref{figent}, it therefore allows one to study the operator complexity in more general settings.

\section{Concluding remarks}
We have explored competition between information scrambling within a system with information leakage to the environment as described by a Markovian master equation. We demonstrated that the Krylov basis, constructed via the bi-Lanczos algorithm, minimises the spread complexity and showed that qualitatively consistent operator dynamics can be captured by considering other suitable bases. Regardless of the specific choice of basis, we established that decoherence caps the size of operators, consistent with earlier results in the thermodynamic limit \cite{liu2024operator}. Our results demonstrate that a basis other than the minimal one can still provide insight into the spread complexity of operator dynamics, opening the possibility to explore the effect of the backflow of information on the competition between scrambling and decoherence. A natural framework for these is using master equations with time-dependent rates~\cite{ryan2022commutativity} or collision models with non-zero Markov order.

\acknowledgments
The authors thank P. Poggi, A. Nico-Katz, and A. Touil for the insightful discussions. EC acknowledges support from the Irish Research Council Project ID No. GOIPG/2020/356 and the Thomas Preston Scholarship. SC acknowledges support from the Science Foundation Ireland Starting Investigator Research Grant “SpeedDemon” No. 18/SIRG/5508 and the Alexander von Humboldt Foundation. S.D. acknowledges support from  the U.S. National Science Foundation under Grant No. DMR-2010127. AK, SC, and SD acknowledge the John Templeton Foundation under Grant No. 62422.
\bibliographystyle{eplbib}
\bibliography{ref}

\end{document}